\documentclass{article}
\usepackage{graphicx} 
\usepackage{siunitx}
\usepackage{listings}
\usepackage{hyperref}
\usepackage{nomencl} 
\usepackage[letterpaper, total={6in, 9in}]{geometry}
\makenomenclature

\title{Enabling GPU Portability into the Numba-JITed Monte Carlo Particle Transport Code MC/DC\footnote{
This is an Accepted Manuscript of an article published by ANS in Proceedings of the International Conference on Mathematics and Computational Methods Applied to Nuclear Science and Engineering (M\&C 2025) on April 27-30 2025, available at: \url{https://doi.org/10.13182/MC25-47142}} \footnote{
Please cite as: J. P. Morgan, B. Cuneo, I. Variansyah, and K. E. Niemeyer. "Enabling GPU Portability into the Numba-JITed Monte Carlo Particle Transport Code MC/DC," in \textit{Proceedings of the International Conference on Mathematics and Computational Methods Applied to Nuclear Science and Engineering (M\&C 2025)}, pp.1934-1943. (2025). Denver, CO, USA. doi: \href{https://doi.org/10.13182/MC25-47142}{10.13182/MC25-47142}.} \footnote{
Presentation available at: \href{https://doi.org/10.5281/zenodo.15330987}{10.5281/zenodo.15330987}
}}

\author{
  Joanna Piper Morgan$^{1,2}$\footnote{CONTACT: morgajoa@oregonstate.edu, joannapipermorgan@gmail.com}
  \and
  Braxton Cuneo$^{1,3}$
  \and
  Ilham Variansyah$^{1,4}$
  \and
  Kyle E. Niemeyer$^{1,2}$
}

\date{%
    \small{
    $^1$School of Mechanical Industrial and Manufacturing Engineering, Oregon State University\\%
    $^2$Center for Exascale Monte Carlo Neutron Transport (CEMeNT)\footnote{https://cement-psaap.github.io/}\\
    $^3$Department of Computer Science, Seattle University\\
    $^4$School of Nuclear Science and Engineering, Oregon State University%
    }
}

\begin{document}

\maketitle

\begin{abstract}
    The Center for Exascale Monte Carlo Neutron Transport is developing Monte Carlo / Dynamic Code (MC/DC) as a portable Monte Carlo neutron transport package for rapid numerical methods exploration on CPU- and GPU-based high-performance computers.
    In this paper, we describe MC/DC's current event-based GPU algorithm as well as the just-in-time (JIT) compilation scheme we use to enable GPU operability on Nvidia and AMD GPUs from MC/DC's Python source.
    To analyze performance, we conduct runtime tests of the C5G7 k-eigenvalue benchmark problem and a continuous-energy infinite pin cell on Nvidia Tesla V100 GPU, AMD MI250X GPU, and the AMD MI300A APU and make comparison to a dual-socket Intel Xeon Sapphire Rapid CPU node.
    We found that for the multi-group C5G7 benchmark problem, we respectively see a 15$\times$, 0.7$\times$, 12$\times$ speedup on a V100, MI250X, and MI300A over 112 Intel Xeon CPU cores.
    For the continuous-energy infinite pin-cell benchmark, we found speedups of 5$\times$, 3$\times$, 4$\times$ on a V100, MI250X, and MI300A, respectively, over the same CPU node.
\end{abstract}

\section{Introduction}\label{sec:1}

The Center for Exascale Monte Carlo Neutron Transport (CEMeNT) develops Monte Carlo / Dynamic Code (MC/DC) to rapidly investigate novel time-dependent neutron transport algorithms and deploy them at scale on modern heterogeneous supercomputers \cite{morgan_monte_2024}.
MC/DC is scripted in Python and uses a novel development strategy for the field that supports executing Monte Carlo transport functions purely in the Python interpreter or compiling to CPUs (x86, PowerPC-64, and ARM-64) \cite{variansyah_mc23_mcdc} and now GPUs (Nvidia and AMD) using the Numba compiler for Python \cite{lam_numba_2015}.
This development strategy is an effort to abstract the specific hardware-target architecture from the numerical methods designer to enable more rapid methods development.
We have previously shown that in single-threaded workloads, Numba provides up to three orders of magnitude of speedup as compared to transport problems run in the Python interpreter.
On the CPU, MC/DC uses a distributed-memory-parallelism-only paradigm provided through mpi4py \cite{dalcin_mpi4py_2021} (which is also currently enabled for multi-GPU computations) for domain replicated or decomposed problems.

As of version 0.11.1, MC/DC supports continuous and multi-group energy neutron transport physics with constructive solid geometry modeling using surface tracking \cite{transport_cement_mcdc_2024}.
It can solve k-eigenvalue problems as well as perform fully dynamic simulations.
It also supports replicable random number generation for CPU and GPU side computations \cite{rngseed}.
In an initial code-to-code performance comparison, MC/DC was found to run about 2.5 times slower than the Shift Monte Carlo code for a simple problem and showed similar scaling up to hundreds of nodes \cite{variansyah_mc23_mcdc}.
MC/DC's CPU development structure has previously been explored~\cite{morgan_monte_2024, variansyah_mc23_mcdc, morgan2022}.
MC/DC's CPU history-based performance is generally between 2 and 10 times slower than that of production Monte Carlo neutron transport applications \cite{variansyah_mc23_mcdc}; nevertheless, not yet highly optimized, MC/DC still has a sizeable room for improvement.

When targeting GPUs, an understanding of the architecture is needed.
GPUs use a single instruction-multiple thread (SIMT) parallelism paradigm \cite{cuda}, where threads are bound together in teams called warps, or thread-blocks, and are required to do the same operations in absolute unison. 
If threads in the same warp need to take different paths in a program (e.g., different if/else branches or iterating loops a different number of times), each path must be executed serially. 
This behavior is called thread divergence.
Threads that do not belong to the currently executing path are turned off so that the end result of the computation is consistent with the control flow logic.
Usually, mitigating thread divergence will result in higher performance of GPU-enabled applications.
For example, consider a warp that has 32 threads, with each thread requiring a different operation.
That warp will have to run each thread individually, thus resulting in serial execution: 32 clock cycles to perform 32 operations.

In this paper, we first discuss the state of the art of GPU operability in production Monte Carlo neutron transport applications and summarize relevant performance characteristics.
We then discuss MC/DC's current event-based transport algorithm for GPUs and describe in depth how MC/DC is compiled to Nvidia and AMD GPUs.
We describe two benchmark problems of interest: a multi-group C5G7 k-eigenvalue benchmark and a continuous-energy, time-dependent infinite pin cell benchmark.
Then, we conduct a performance analysis using Intel Sapphire Rapids CPUs, Nvidia Tesla V100, and AMD MI250X GPUs, as well as the new AMD MI300A APU on supercomputers at Lawrence Livermore National Laboratory.

\section{Modeling Strategy}
\label{sec:modeling_strat}

In this section, we describe the GPU modeling strategy of several state-of-the-art production codes---namely Shift, OpenMC, and Mercury---and then describe MC/DC's event-based algorithm.
Shift is a Monte Carlo neutron transport package available as part of the SCALE library from Oak Ridge National Laboratory \cite{pandya_implementation_2016}.
Shift allows for multi-group and continuous-energy transport on Nvidia and AMD GPUs.
The Shift development team has reported that, for multi-group GPU algorithms, all GPUs on a given node allow for between 7 and 80 times speedup compared to all CPUs of those same nodes for a set of C5G7-type problems \cite{hamilton_multigroup_2018}.
The Shift team has also reported that, for most problems, the performance on an Nvidia Tesla V100 is about equal to that on one graphics compute die of an AMD MI250X, which in turn is equivalent to about 100--150 Intel Xeon CPU cores \cite{mcsummit}.

Similarly, GPU support for the Mercury Monte Carlo neutron transport solver from Lawrence Livermore National Laboratory is enabled by an event-based algorithm written in C-CUDA \cite{pozulp_progress_2023}.
The Mercury development team has discussed the importance of optimizing code generation through the use of language/compiler tools such as link time optimization (LTO) and whole code optimization.
For a Godiva in water continuous-energy benchmark problem, Mercury has shown up to a 7.7 times speedup on a whole node of the Sierra class systems as compared to a whole node of Intel Xeon dual-socket type CPUs \cite{pozulp_progress_2023}.
The Mercury development team has announced initial development efforts porting to AMD MI300A APUs \cite{pozulp_sna_2024}.
Other GPU codes have been developed as standalone C-CUDA projects directly targeting GPUs, including PRAGMA \cite{choi_optimization_2021}, GUARDYAN \cite{molnar_gpu_based_2019}, and a CUDA-based port of OpenMC \cite{ridley2021}.

OpenMC from Argonne National Laboratory differs from these approaches.
OpenMC uses the OpenMP target-offloading model to compile and execute on Nvidia, AMD, and Intel GPUs for event-based transport algorithms.
OpenMC is currently the only Monte Carlo neutron transport code enabled to run on Intel's PVC GPUs.
Furthermore, the OpenMC team has demonstrated node scaling on Nvidia, AMD, and Intel GPUs.
The OpenMC development team reports that, for the continuous-energy exaSMR test problem, they achieve a 70 times speed up on Intel GPUs as compared to a dual-socket Intel Xeon CPU node \cite{tramm2024performanceportablemontecarlo}.

When targeting GPUs, MC/DC uses the Harmonize GPU runtime to orchestrate execution and data marshaling, decreasing thread divergence and increasing performance on GPUs \cite{brax2023}.
In this mode, Numba is used to compile transport functions to intermediate compiler representations (PTX or LLVM-IR for Nvidia and AMD hardware-targets, respectively), which in turn are bundled with device, global, and host functions that promote optimal runtime behavior of functions from Harmonize (more on this in Section~\ref{sec:jit}).
This abstraction technique allows for GPU compilation/execution to be treated generically and distinctly from numerical methods development and Monte Carlo transport logic.
Thus, the GPU architecture is largely abstracted from the MC/DC developer, who works entirely in Python script.
The hope is that this will allow for both portability to various GPU types and also enable rapid numerical methods development for time-dependent Monte Carlo transport algorithms.

Harmonize has two scheduling strategies: an event-based and an asynchronous algorithm.
The asynchronous algorithm is described in depth by Cuneo and Bailey~\cite{brax2023} as well as an additional publication in this conference proceeding.
Asynchronous scheduling is currently only available for Nvidia GPUs; thus, in this paper, we focus on the more generic event-based execution strategy which is currently deployed on AMD and Nvidia GPUs.

MC/DC's initial implementation of event-based transport uses a monolithic event design.
In this mode all transport operations (e.g. sample direction, move particle location, etc.) are are ran together as single event (also referred to as a single segment) of transport.
No pre-kernel launch sorting or grouping of like operations is currently provided.
This execution structure is similar to what Hamilton, Slattery, and Evans call \textit{history-length truncation} \cite{hamilton_multigroup_2018} if the truncation criterion was set to one.

For a GPU-centric perspective, say a warp of 32 threads receives 32 live particles from the stack to transport.
The physical ``events'' that those particles require could be completely different.
If there are indeed 32 different events then the warp will execute serially in order to process the single step (or segment) for those 32 particles.
Better occupancy (how many threads are able to operate in unison) may be expected for problems where singular events are dominant (e.g., highly scattering problems).

Furthermore, modern GPUs may concurrently execute many warps within the same streaming multiprocessor (SM), allowing the SM to hide the latency of loads by switching from stalled warps to eligible ones.
This monolithic event structure is paired with a dual set of particle banks where one stores particles that have yet to be processed and the second stores those that have already been processed by the executing kernel.
As the monolithic scheme does not subdivide the logic of the particle transport loop, there may be no savings in register (high-speed warp memory) allocation that is normally enjoyed when implementing an event-based transport algorithm on GPUs.


Work is ongoing to move to a more field-standard event-based approach where the main transport operations are decomposed to be a more narrowly defined set of operations where each event is individually launched \cite{brown_stack}.
This alteration will make our event-based transport algorithm more in line with what is implemented in GPU-enabled production codes \cite{hamilton_multigroup_2018, pozulp_progress_2023, choi_optimization_2021, molnar_gpu_based_2019, tramm2024performanceportablemontecarlo}.

\section{JIT Compilation process}
\label{sec:jit}

When targeting GPUs, MC/DC functions are just-in-time (JIT) compiled with Harmonize.
To JIT compile and execute on AMD or Nvidia GPUs, MC/DC users need only to append their terminal launches with a \texttt{--target=gpu} option.
When considered in totality the MC/DC+Numba+Harmonize JIT compilation structure is akin to ``portability framework", in that it allows dynamic targeting and developer abstraction of hardware architectures, like OpenMP target-offloading used by OpenMC.
This JIT compilation process allows MC/DC to pair the idea of a portability framework with a high-level language in an effort to enable more rapid methods development on Exascale systems.

Monte Carlo transport functions from MC/DC are treated as device functions with global, host, and additional device functions coming from Harmonize.
Mixing codes from various sources (Python and C++) requires the user to provide an \textit{exacting} set of compiler options to achieve an operable executable.
We provide in-depth descriptions of these sets of commands as we found the definition of this JIT compilation process one of the most difficult parts to get the MC/DC+Harmonize software engineering structure operable.

\begin{figure}[]
  \centering
  \includegraphics[width=\textwidth]{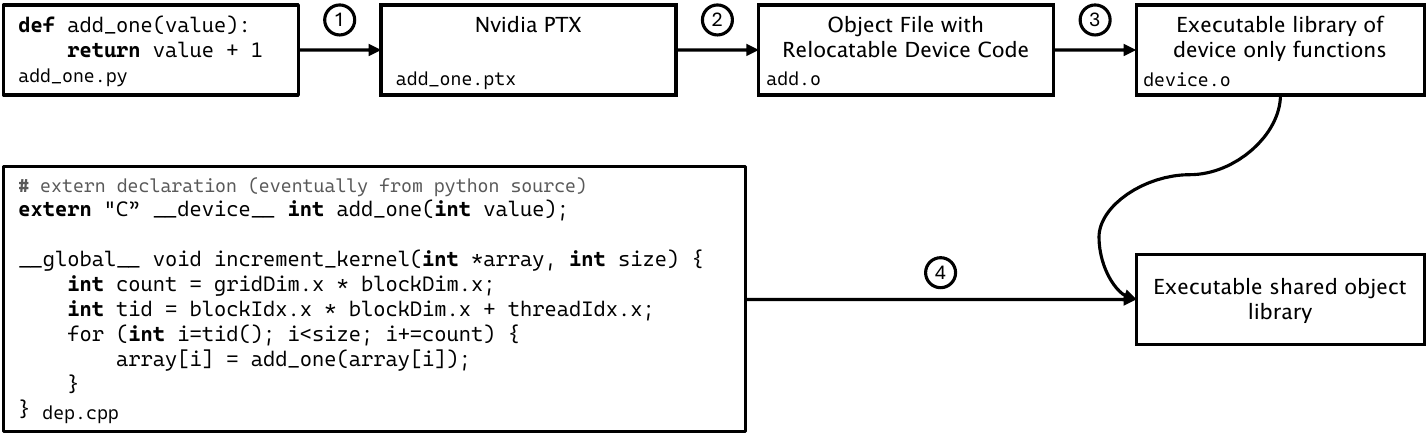}
  \caption{Simple proxy example describing how to compile device functions in Numba-Python with external C++ code for targeting Nvidia GPUs. In this simplified proxy, the Python function corresponds to MC/DC, and the C++ code corresponds to Harmonize.}
  \label{fig:codenvcc}
\end{figure}

To examine the compilation strategy in-depth, a simple proxy problem is provided in Figures~\ref{fig:codenvcc} and \ref{fig:codeclang}.
The figures show a simple Python function that does integer addition on a provided value (representing MC/DC transport operations) and a C++ snippet (representing Harmonize) showing first the declaration of an extern device function (eventually coming from Python) and a global function which will act as the GPU runtime for our Python device function.
Note that for the operability of these examples, extra functions are required in \texttt{dep.cpp} and \texttt{add\_one.py} but are truncated for brevity.

\subsection{Nvidia Targets}

To compile to Nvidia GPU hardware-targets at runtime, we rely entirely on the Nvidia C-Compiler (\texttt{nvcc}).
Current versions of Numba come with CUDA operability natively, but this is set to be deprecated in future releases in favor of a more modular approach where the Numba-CUDA package will be an optional separate feature.

We begin by
\begin{enumerate}
    \item Compiling Python device code to Nvidia PTX by \texttt{numba.cuda.compile\_ptx\_for\_current\\\_device} (which requires typed function signatures), then place that output into \texttt{add\_one.ptx} file; next
    
    \item Compiling PTX to relocatable device code using \texttt{nvcc -rdc=true -dc -arch=<arch> --cudart shared --compiler-options -fPIC add.ptx -o add.o} where \texttt{-dc} asks the compiler for device code, \texttt{-rdc} asks to make that device code relocatable, \texttt{--cudart shared} asks for shared CUDA runtime libraries and \texttt{-fPIC} generates position-independent code;
    
    \item Compiling that relocatable byte code into a library of executable device functions is done with \texttt{nvcc -dlink add.o -arch=<arch> --cudart shared -o device.o --compiler-options \\-fPIC} where \texttt{-dlink} asks the compiler for relocatable device code; and finally
    
    \item Compiling the C-CUDA file containing the global function and linking with the library of device functions originating from Python with \texttt{nvcc -shared add.o device.o -arch=<arch> --cudart shared}.
    
\end{enumerate}

While the complexity of the functions both from MC/DC (Python) and Harmonize (C++) increases dramatically when moving toward implementation in MC/DC, this compilation strategy remains mostly the same.
The exact compilation commands Harmonize calls when compiling MC/DC functions can be viewed by setting \texttt{VERBOSE=True} in \texttt{harmonize/python/config.py}.
This compilation strategy also allows for the extension of functions defined in the CUDA API but not in Numba-CUDA as they can come from the C-CUDA source in \texttt{dep.cpp}.

\subsection{AMD Targets}
Just in time compilation and execution to AMD devices are enabled as of MC/DC v0.11.0 \cite{transport_cement_mcdc_2024}.
Significant adaptations from the process of Nvidia compilation are required to target AMD GPUs.
PTX is a proprietary Nvidia standard, so when targeting AMD GPUs, we rely on intermediate compiler representation (IR) from LLVM for an AMD GPU hardware-target (also called an LLVM target triple).
AMD's compiler toolchain is based in the LLVM-Clang ecosystem, so we will be calling LLVM-Clang-based tools (e.g., \texttt{hipcc} is a wrapper function for \texttt{clang}).
Note that while the LLVM-Clang commands are generic, AMD variations of compilers, linkers, etc. must be invoked.
For example, to invoke the correct Clang compiler point to the ROCm installed variation (often on LinuxOS at \texttt{opt/rocm/llvm/bin/clang}).

To generate AMD target LLVM-IR from Python script, a patch to Numba is provided by AMD\footnote{https://github.com/ROCm/numba-hip}.
This patch can also execute produced functions from the Python interpreter, much like Numba-CUDA.
As this patch is a port of AMD's Heterogeneous-computing Interface for Portability (HIP) API, it attempts to be a one-to-one implementation of operations implemented in Numba-CUDA.
The Numba-HIP development team has gone as far as to provide a \texttt{numba.hip.pose\_as\_cuda()} function, which, after being called in Python script, will alias all supported Numba-CUDA functions to Numba-HIP ones and compile/run automatically.

\begin{figure}[]
  \centering
  \includegraphics[width=\textwidth]{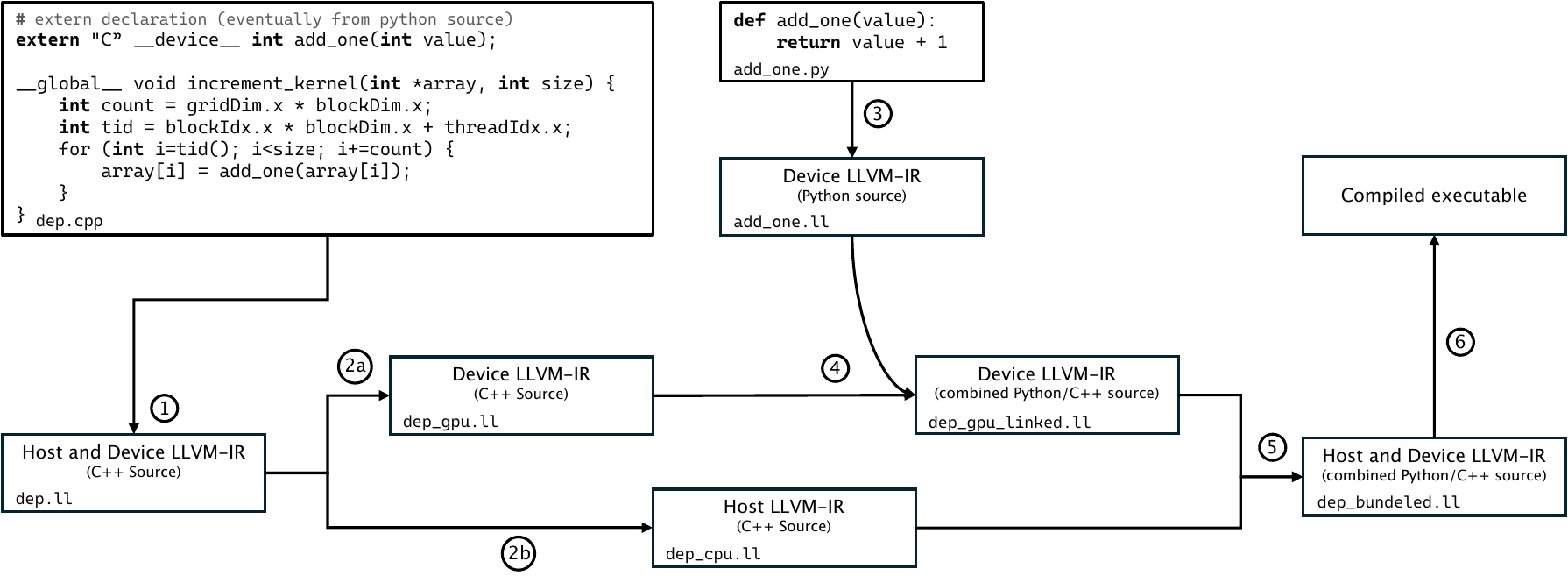}
  \caption{Simple proxy example describing how to compile device functions in Numba-HIP with external C++ code to AMD GPU targets. In this simplified proxy, the Python function corresponds to MC/DC, and the C++ code corresponds to Harmonize}
  \label{fig:codeclang}
\end{figure}

When moving to compile and execute full MC/DC+Harmonize, we must again enable the compilation of device functions from Numba-HIP and device, global, and host functions from C++.
To show that process, we again explore a simple proxy application shown in figure \ref{fig:codeclang} where a Numba-HIP function adds one to an integer value and a C++ function declares an extern function by the same name and runs that function for all values of an array.

Every GPU program is technically a bound set of two complementary applications: one that runs on the host side (CPU) and the other on the device side (GPU), with global functions linking them together.
To link external device code together for AMD hardware-targets, we have to unbundle these two programs, link the extra device functions (coming from Python) to the device side, then re-bundle the device and host functions back together.
This process is done in LLVM-IR.

Figure \ref{fig:codeclang} shows the compilation structure.
We begin compilation by
\begin{enumerate}
    \item Compiling C++ source in \texttt{dep.cpp} to LLVM-IR with host and device code bundled together with \texttt{hipcc -c -fgpu-rdc -S -emit-llvm -o dep.ll -x hip dep.cpp -g} where \texttt{-fgpu-rdc} \\ asks the compiler for relocatable device code \texttt{-emit-llvm} requests the LLVM-IR, \texttt{-c} only runs preprocess, compile, and assemble steps, and \texttt{-x hip} specifies that \texttt{dep.cpp} is HIP code;
    
    \item Unbundling the LLVM-IR:
    
    \begin{enumerate}
        \item first the device half \texttt{clang-offload-bundler --type=ll --unbundle --input=dep.ll \\--output=dep\_gpu.ll --targets=hip-amdgcn-amd-amdhsa--gfx90a} \\ where \texttt{amdgcn-amd-amdhsa} is the LLVM target-tipple and \texttt{gfx90a} is compiler designation for an MI250X;
        
        \item then the host half \texttt{clang-offload-bundler --type=ll --unbundle --input=dep.ll --output=dep\_cpu.ll --targets=host-x86\_64-unknown-linux-gnu}; then
    \end{enumerate}
    
    \item Compiling device functions from Python source with \texttt{numba.hip.generate\_llvmir()} and place into \texttt{add\_one.ll};
    
    \item Linking the now unbundled device code in \texttt{dep\_gpu.ll} and the device code from Python in \texttt{add\_one.ll} together with \texttt{llvm-link dep\_gpu.ll add\_one.ll -S -o dep\_gpu\_linked.ll};
    
    \item Rebundling the now combined Python/C++ device LLVM-IR back to the host LLVM-IR with \texttt{clang-offload-bundler --type=ll --input=dep\_gpu\_linked.ll --input=dep\_cpu.ll \\--output=dep\_bundled.ll --targets=hip-amdgcn-amd-amdhsa--gfx90a,\\host-x86\_64-unknown-linux-gnu}; and finally
    
    \item Compiling to an executable with \texttt{hipcc -v -fgpu-rdc --hip-link dep\_bundled.ll\\-o program} where \texttt{--hip-link} links clang-offload-bundles for HIP
    
\end{enumerate}

As in the Nvidia compilation, non-implemented functions can be brought into the final program via the C++ source.
This was required for MC/DC on AMD GPUs as vector operable atomics are not currently implemented in the Numba HIP port and thus must come from the C++ side.
We hope that these more generic adaptations (relying on LLVM-Clang infrastructure instead of CUDA) will allow for greater extensibility as we move to target future accelerator platforms---namely, Intel GPUs.
For compilation to Nvidia hardware-targets, we will still keep the PTX-based compilation structure.

\section{GPU performance analysis}

To examine the performance of MC/DC+Harmonize's monolithic event and JIT compilation strategy, we run two problems on HPC nodes and compare the runtime performance between all the GPUs of a given node and a dual-socket Intel Xeon Sapphire Rapids node.
To compare, we use a CPU core equivalency equation
\begin{equation}
    \text{Effective CPU cores per GPU} = \frac{\text{\#CPU cores}\times\text{CPU run time}}{\text{\#GPUs}\times\text{GPU run time}} \; ,
\end{equation}
to have a core equivalency measure on a per GPU basis.

We first examine a fully 3D C5G7 benchmark problem with multi-group cross-sections and geometries as defined by Hou et al.~\cite{hou2017}.
We run a k-eigenvalue simulation with 50 inactive and 100 active cycles with \num{1e6} particle histories modeled in every iteration.

\begin{figure}[]
  \centering
  \includegraphics[width=\textwidth]{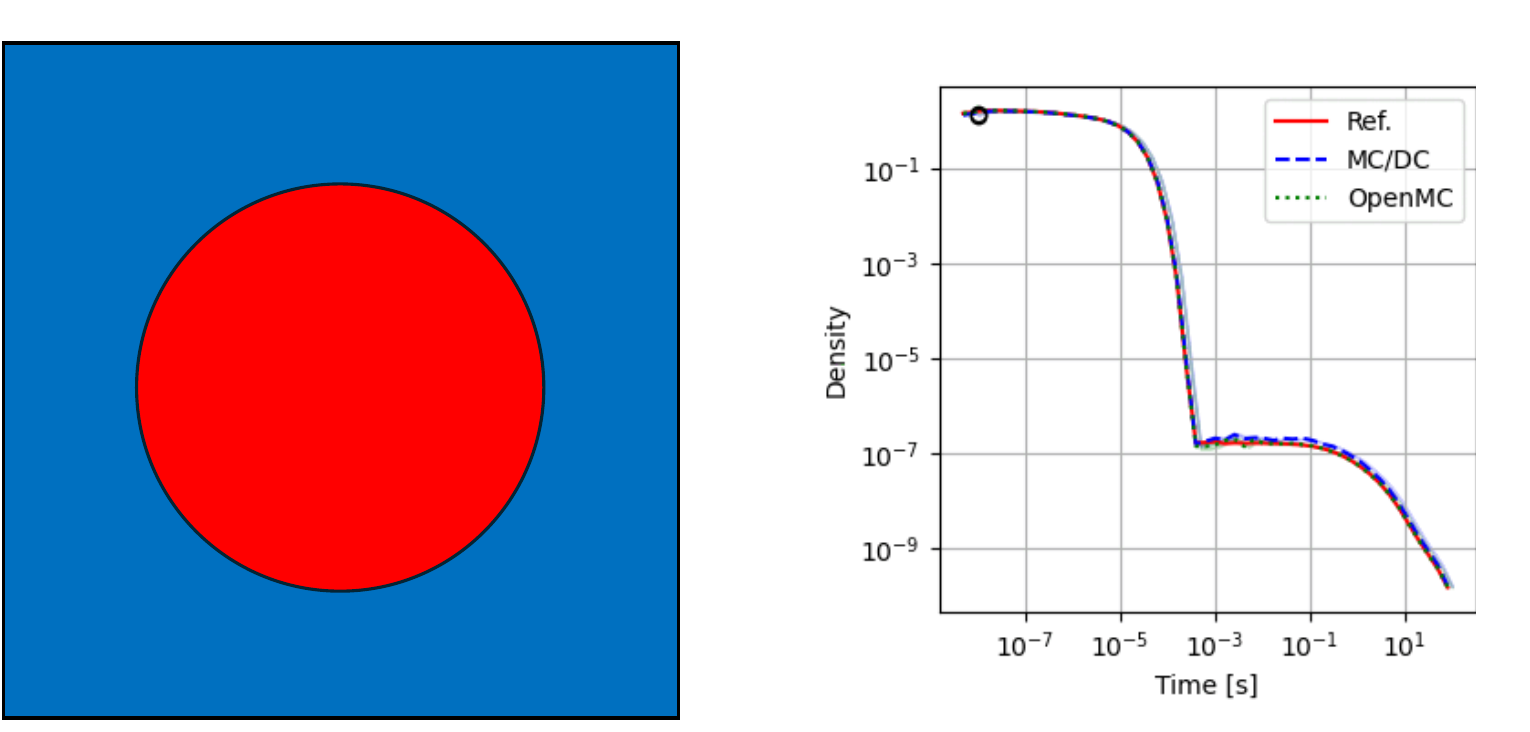}
  \caption{Left: Infinite pin cell with reflecting boundaries on all sides, Red is 2.4\% enriched UO$_2$, and blue is a homogeneous Boron/Water moderator. Right: Neutron population density of infinite pin cell benchmark problem through time from MC/DC and OpenMC at \num{1e7} particle histories and a reference from OpenMC at \num{1e9}}
  \label{fig:inf_pin}
\end{figure}

Next, we examine the performance of an infinite Uranium dioxide (UO$_{\text{2}}$) fuel pin at 2.4\% enrichment surrounded by a homogeneous mixture of water and Boron shown in Figure~\ref{fig:inf_pin} at left.
A uniform 14 MeV source drives the problem and is molded as a continuous-energy problem with cross-sections from an OpenMC's Official Data Library\cite{romano_openmc_nodate}.
To aggressively verify MC/DC's continuous-energy physics, we compare neutron density through time produced from MC/DC and OpenMC in figure \ref{fig:inf_pin} at right.
Both MC/DC and OpenMC solutions are produced with \num{1e7} particle histories, and a reference solution from OpenMC is also shown produced from \num{1e9} particle histories.
MC/DC's solution does deviate from OpenMC's and the reference solutions as MC/DC currently has several limitations when running in continuous-energy mode.
Notably, $S(\alpha, \beta)$ scattering and high energy reactions are not currently implemented, and we also assume the free-gas scattering model with constant cross-sections.
Taking this into account, there is good agreement between the flux spectrum evolution of the three solutions.
The full problem definition for the continuous-energy problem is provided in \texttt{examples/fixed\_source/inf\_pin\_ce/input.py} in v0.11.1 of MC/DC's
repository \cite{transport_cement_mcdc_2024}.
To conduct performance analysis, we use 10$^7$ particle histories run in time-dependent fixed-source mode.

We ran these problems on high-performance computing systems available at Lawrence Livermore National Laboratory (LLNL): the Dane, Tioga, and Lassen machines.
Dane is a CPU-only system with dual-socket Intel Xeon Sapphire Rapids CPUs, each with 56 cores for a total of 112 per node. 
We will make all CPU-GPU performance comparisons against CPU runtime from a whole node (112 threads) of Dane.
Lassen is the open collaboration sibling to the Sierra machine with four Nvidia Tesla V100s and two IBM Power 9 CPUs per node.
Tioga is an early access machine for LLNL's exascale-class El Capitan machine.
On its standard partition, Tioga's nodes have four AMD MI250X GPUs and one AMD EPYC 7A53 CPU.
Note that a single MI250X itself has two graphics compute dies (GCDs), so one node of Tioga has a total of eight GPUs for MPI purposes; we still report results as four GPUs (as is standard for this hardware).
Tioga also has a partition of AMD MI300A APUs, which will be deployed in the exascale-class El Capitan machine.
Each MI300A is a 24-core CPU on the same die as the GPU.
The APU architecture of the MI300A is different from previous generations of GPUs as there is little to no overhead cost to moving data from the host and device side as device memory is shared between the CPU and GPU components of the APU.

\section{Results, Discussion, and Future Work}

We compile to AMD and Nvidia GPUs with ROCm v6.0.0 (clang v17.0.0) and CUDA v11.8, respectively, 
and to CPU and Nvida-PTX with Numba v0.58.1.
For AMD GPU targeted LLVM-IR, AMD provides the Numba HIP patch, which we apply to Numba v0.58.1. 
We run MC/DC v0.11.1 \cite{transport_cement_mcdc_2024} and Harmonize v0.0.2 \cite{harmonize}.
We use double precision for all floating point operations.

\begin{table}[ht]
    \centering
    \caption{Run time and CPU core equivalent for C5G7 (multi-group) and infinite pin-cell (continuous-energy) benchmark problems from various hardware types.}
    \begin{tabular}{@{}c c c c c@{}}
        \hline
        (\#) Hardware & C5G7 & & Infinite Pin-cell & \\
        & Runtime [s] & Core Equivalency & Runtime [s] & Core Equivalency \\
        \hline
        (112) Sapphire Rapids CPU & 5080. & $-$ & 512.4 & $-$ \\
        (4) Nvidia V100 & 342.6 & 415 & 111.0 & 129 \\
        (4) AMD MI250X & 7020. & 20 & 181.2 & 79 \\
        (4) AMD MI300A & 436.8 & 326 & 116.4 & 123 \\
        \hline
    \end{tabular}
    \label{tab:runtime}
\end{table}

Table \ref{tab:runtime} shows runtime and CPU core equivalencies for the multi-group C5G7 and continuous-energy infinite pin-cell benchmark problems.
For the C5G7 core equivalences, V100 and MI300A both perform well as compared to Intel CPUs, with V100s producing a 15$\times$ speedup corresponding to a 415 CPU core equivalency per GPU.
Similarly, MI300A sees a 12 $\times$ speedup corresponding to 362 CPU cores per GPU.
Regardless, we see an order of magnitude speedup using a whole node of these two GPU types compared to the whole node of a dual-socket Intel Xeon node which is similar to what has been reported from other GPU-based codes \cite{tramm2024performanceportablemontecarlo, mcsummit}.
MI250X performance shows a slowdown relative to CPU performance.
Performance analysis using ROCm profilers available on Tioga is underway in an effort to explanation for this slowdown.

Moving to the infinite pin-cell continuous-energy problem, MI250X performance is now more in line with V100 GPUs and MI300A APUs. Core equivalency is clustered between 80 to 130, and the speedup is 5$\times$, 3$\times$, and 4$\times$ for V100s, MI250X, and MI300A, respectively. 
We believe this decrease in relative performance between CPU and GPU is due to the monolithic event kernel, specifically when handling the large number of cross-section lookups and interpolations that promote thread divergence.
Again, V100s seem to be the most performant (with core equivalency at 129) but only slightly above the MI300A (at 123).
MI300A outperforms MI250X significantly for the multigroup C5G7 eigenvalue problem but only slightly for the continuous-energy time-dependent pin-cell problem. 

More work is needed to do further code-to-code comparisons of MC/DC to both production CPU and GPU codes.
We believe when this is coupled with profiling, particular deficiencies with our algorithms will be identified.
When these are addressed (most likely using similar algorithms implemented by production Monte Carlo transport codes on GPU), we will see performance improvements on all GPU hardware-targets.

The Shift development team has reported that one GCD of an MI250X is roughly equal to the performance of one V100 \cite{mcsummit}.
We are currently not able to replicate this result in MC/DC.
We believe this is, at least partly, due to non-optimal code generation for AMD hardware-targets.
The Mercury development team has previously shown how important compiler optimizers are in the production of performant GPU code \cite{pozulp_progress_2023}.
Compilation times may loosely indicate how much automatic compiler optimization is occurring.
For the C5G7 problem, compiling to Nvidia GPU targets took more than three times as long (at about \SI{160}{s}) as compiling to AMD GPU targets (about \SI{45}{s}).
This could indicate a lack of automatic compiler code optimization, which may be caused by our JIT-compilation interrupting ROCm's compilation path.
We will experiment with additional compiler flags to force more optimization when generating device code.


Work is ongoing to improve the GPU and CPU performance of MC/DC.
The most important future analysis will be the use of profilers to identify GPU code hot spots and address them with either algorithmic or compiler optimizations.
Eventually, we will run the continuous-energy exaSMR problem so we can conduct a direct comparison with other production Monte Carlo codes on GPU nodes.
We will also begin exploring node-scaling for problems of interest.
Furthermore, we will continue to engage in novel method developments for transient problems of interest.

\section*{ACKNOWLEDGEMENTS}
The authors would like to thank Dr. Damon McDougall and Dr. Dominic Etienne Charrier from AMD for support with Numba-HIP and ROCm compilers.
The authors would also like to thank the Livermore computing staff for continued support using the Dane, Tioga, and Lassen machines and Dr. Steven Hamilton from Oak Ridge National Laboratory for supplying alternate versions of C5G7 cross-section data and geometries.
This work was supported by the Center for Exascale Monte-Carlo Neutron Transport (CEMeNT) a PSAAP-III project funded by the Department of Energy, grant number: DE-NA003967.

\bibliographystyle{IEEEtran}
\bibliography{main}

\end{document}